\documentstyle[psfig,subfigure,lscape]{mn}

\begin{document}
\newcommand{\goodgap}{%
 \hspace{\subfigtopskip}%
 \hspace{\subfigbottomskip}}

\title[SCUBA observations of MAMBO sources]
{SCUBA observations of MAMBO sources}
\author[S. Eales et al.]{Steve Eales$^{1}$\thanks{E-mail:
sae@astro.cf.ac.uk}, Frank Bertoldi$^2$, Rob Ivison$^3$, Chris Carilli$^4$,
\newauthor
Loretta Dunne$^1$
and Frazer Owen$^4$\\
$^{1}$Department of Physics and Astronomy, 
Cardiff University, P.O. Box 913, Cardiff CF2 3YB, UK \\
$^{2}$ Max-Planck-Institut f\"ur Radioastronomie, Auf dem H\"ugel 69,
53121 Bonn, Germany\\
$^{3}$ Astronomy Technology Centre, Royal Observatory, 
Blackford Hill, Edinburgh EH9 3HJ, UK\\
$^{4}$ NRAO, Socorro, NM 87801, USA}

\maketitle                  

\begin{abstract}
We have observed 23 sources from the MAMBO 1200$\mu$m survey with
SCUBA at 850$\mu$m, detecting 19 of the sources.  
The sources generally have low values for the ratio of
850$\mu$m to 1200$\mu$m flux. Two possible explanations
for the low values are either that the sources are at very high redshifts
or that the global properties of the dust in the MAMBO sources are
different from the global properties of dust in 
low-redshift galaxies. If the former explanation is
correct,
we estimate that 15 of the MAMBO sources lie at
$\rm z > 3$.
\end{abstract}

\begin{keywords}
submillimetre-dust-galaxies:evolution-galaxies:formation
\end{keywords}

\section{Introduction}       
The luminous high-redshift dust sources discovered by the
SCUBA submillimetre and MAMBO millimetre surveys 
(Smail, Ivison \& Blain 1997; Hughes
et al. 1998; Barger et al. 1998;
Eales et al. 1999; Bertoldi et al.
2000,2001) are 
of great significance for our understanding of
galaxy formation. 
The ultimate energy source in these objects
is hidden by dust but the two
obvious possibilities are that (1) the dust is being heated by a hidden active
nucleus or (2) the dust is being heated by a luminous population of stars.
The first of these possibilities can now largely be ruled out
because of the  
failure of the XMM/Newton and
Chandra telescopes
to detect strong X-ray emission from 
many of the dust sources (e.g. Almaini et al. 2003;
Waskett et al. 2003). Estimates of the star-formation rates necessary
to produce the dust luminosity can be as high as $\rm 6 \times 10^3\ M_{\odot}$
(Smail et al. 2003), enough to produce the stellar population of a massive
galaxy in $\sim 10^8-10^9$ years. 
Many authors have concluded that these dust sources
are the ancestors of present-day elliptical galaxies, basing their
arguments on estimates of the star-formation rate in the population
as a whole (Smail, Ivison and Blain 1997; Hughes et al. 1998;
Blain et al. 1999),
on estimates of the contribution of the sources to the extragalactic background
radiation (Eales et al. 1999), and on comparisons of the space-density
of the SCUBA/MAMBO sources (henceforth SMS) 
with the space-density of ellipticals in the
universe today (Scott et al. 2002; 
Dunne, Eales and Edmunds 2003). 

We are still, however, remarkably ignorant about this
population.
A major problem has been the lack of accurate redshifts for the
SMSs. The practical difficulties here are the large
errors on the positions of the SMSs, which often make
it difficult to determine
the optical/IR counterpart to the SMS,
and the faintness of these counterparts, which make it
difficult to measure an optical spectroscopic redshift.
Until recently, the recourse of most groups has been to
{\it estimate} the redshifts from the ratio of
radio to submillimetre flux.
The surface density of radio sources in deep VLA radio surveys is
low enough that it is possible to be confident that apparent radio
counterparts to SMSs are not chance coincidences.
About 50\% of the SMSs are also faint radio sources
(Smail et al. 2000; Ivison et al. 2002; Clements et al. 2003). The discovery
that a large fraction of SMSs are radio sources is extremely useful,
because for these SMSs it
is then possible, because of the accurate radio positions,
to determine 
the optical/IR counterpart to the SMS. Furthermore, Carilli and Yun (1999)
pointed out that, if SMSs are star-forming galaxies like those in the
universe today, it is possible to estimate the redshift of
the SMS from the ratio of radio to submillimetre flux.

Recently, however, Chapman et al. (2003) have taken a major step forward
by measuring the redshifts for a significant number of SMSs.
Rather surprisingly, given the dust in these objects, this group
succeeded in detecting Lyman $\alpha$ and other $UV$ lines with the
Keck Telescope in 10 radio-selected SMSs. The redshifts they have
measured lie in the range $\rm 0.8 < z < 4$ with a interquartile
range of $\rm 1.9 < z < 2.8$, although this distribution
may be skewed towards low redshifts because of the requirement that
the SMSs be detected at radio wavelengths.

A less widely-appreciated problem is the unknown temperature
of the dust in the
SMSs.
The strong dependence of
bolometric luminosity on dust temperature
means that
estimates of the star-formation rates, both of individual
objects and of the population as a whole, are extremely uncertain.
To our knowledge, there is no SMS without a powerful active
nucleus
which has a sufficiently well-sampled spectral energy distribution 
to permit an accurate estimate of the dust temperature. 
This is a fundamental problem because although the SMSs are often assumed
to be similar to the Ultraluminous Infrared Galaxies in the nearby universe,
the constraints on the spectral energy distributions are consistent
with the SMSs being much colder and less luminous than ULIRGs (Eales
et al. 2000; Efstathiou and Rowan-Robinson
2003). These first two problems are connected. Because
of the degeneracy between dust temperature and redshift
(Blain 1999), it is impossible to make an accurate estimate
of the dust temperature without first measuring the redshift.

A third problem is simply the uncertain reliability of submillimetre
and millimetre surveys. The efficiency of submillimetre
surveys with SCUBA and MAMBO is extremely low, requiring typically eight hours
of observations to produce a single SMS. Something that is again not
widely appreciated is that if these instruments
had been only a factor of two
less sensitive virtually no SMSs would have been detected. These facts
mean that most SMSs are quite close to the detection limits of the
surveys, and low signal-to-noise combined with highly confused fields
lead to interesting problems in data analysis. Different groups have adopted
different data-reduction procedures, and there has been much vigorous debate
within the community about the reliabililty of the different procedures
and about the fraction of SMSs which are likely to be genuine. 

In this paper we present SCUBA observations at 850$\mu$m 
of a sample of sources from a survey with MAMBO at 1200$\mu$m.
There were two motives for our programme. The first was to
provide a simple cross-check between the surveys. Are the SMSs detected
by MAMBO confirmed by our SCUBA observations? Are the ratios of
850$\mu$m to 1200$\mu$m flux what one would expect for
star-forming galaxies at high redshift? 
The second motive was
to use this ratio to determine whether there is a population
of SMSs at very high redshift; 
at high redshifts this ratio falls as the wavelengths probed
move, in the rest frame of the source, towards the peak 
of the dust spectral energy distribution.
We assume a value for the Hubble constant of
$\rm 75\ km\ s^{-1}\ Mpc^{-1}$.

\section{The MAMBO Sample}

The Max-Planck Millimeter Bolometer array
(MAMBO, Kreysa et al. 1998)
on the IRAM 30m telescope
has been used to carry out the first
survey of the sky at millimetre wavelengths.
The survey is being carried out in three main fields: the 
Lockman Hole, the NTT
Deep Field, and a field centred on the cluster Abell 2125. The survey has
already
discovered roughly the same number of dust sources as have been
found in the surveys
with SCUBA at 850$\mu$m. Some preliminary results from the survey
were presented by Bertoldi et al. (2000,2001) and Dannebauer et al. (2002).

We produced a sample of MAMBO sources to observe with SCUBA 
by selecting all sources in the MAMBO
catalogue (as of mid 2001) with S/N greater than four and then
imposing a lower flux limit of 3.3 mJy in the NTT field and 3 mJy
in the other two fields. In the NTT field there are eight sources which met
these criteria. We observed seven of these with SCUBA
and also one other source (NTT-MM22)
which falls slightly
below our flux limit. In the Abell 2125 field there are
19 sources which met our criteria. We observed ten of these
and also one other source (Abell2125-MM50) which falls slightly
below our flux limit. In the Lockman Hole field there are six sources
which met our selection criteria. We observed two of these sources, and
two other sources (LH-MM34 and LH-MM8)
fall within the area surveyed at 850$\mu$m
as part of the SCUBA 8mJy survey (Scott et al. 2002).

Although we have observed almost all the sources in the
NTT sample, we have failed to observe (through pressure
of observing time) a significant fraction
of the sources in the other two samples. The sample of
sources which we did observe is biased in one important respect.
Our SCUBA observations (\S3) were made in photometry mode, in which
a single bolometer is pointed at the target position. For these
observations to be reliable, accurate positions are clearly
essential. 
We therefore gave precedence to sources with accurate positions,
either from observations
with the Plateau de Bure millimetre interferometer
(Dannebauer et al. 2002) or from VLA radio observations.
Our bias towards sources with VLA detections (only one of the sources
in our statistically-complete MAMBO samples which we did not observe
with SCUBA was definitely detected by the VLA) creates a potential
redshift bias in our project.
The 
ratio of radio to millimetre
flux is expected to decrease with redshift (Carilli and Yun 1999),
and so the sample of MAMBO sources we observed with SCUBA may be biased
towards low redshifts. We estimate, from the predicted relationship 
between the
radio/submillimetre ratio and redshift (\S 6), that the
MAMBO sources which were not detected by the VLA lie mostly
at $\rm z > 3$.

Table 1 lists all the sources we observed.
The position given for
each source is the position we used for the SCUBA photometry. In order
of preference, we used the Plateau de Bure position, the VLA position
or the original position found in the MAMBO survey. For the sources
for which we used the VLA position, we have also given our estimate
of the probability that the VLA source is not actually associated with
the MAMBO source. This is given by $p = 1 - exp(-n \pi d^2)$, in which
$d$ is the angular distance between the MAMBO source and radio source and
$n$ is the surface density of radio sources on the sky, which
we 
calculated using the 1.4 GHz source
counts given in Richards (2000).

\begin{table*}
\(\begin{tabular}{|l|l|l|l|l|l|l|l}
\multicolumn{5}{|c|}{Table 1. The Sample}\\
\hline
(1)&(2)&(3)&(4)&(5)&(6)&(7)&(8)\\

Name & $\rm S_{1200\mu m}/mJy$ & Position (2000.0) 
& Type & 1.4-GHz flux/$\mu$Jy & Radio ref.& d/arcsec & P(radio) \\
\hline

LH-MM13 & $4.2\pm1.0$ & 10 52 01.05 57 24 46.10 & R & $\rm 73\pm10$ & I & 7.6& 0.055 \\
LH-MM86 & $4.6\pm0.6$ & 10 52 14.18 57 33 28.3 & R & $\rm 74\pm8$ & B & 1.0 & $\rm 9.6\times10^{-4}$ \\

\hline

NTT-MM3 & $4.6\pm1.0$ & 12 05 08.13  -07 48 11.8 & R & $\rm 88\pm15$ & B & 2.1 & 0.0043 \\
NTT-MM34 & $3.3\pm0.7$ & 12 05 09.75 -07 40 02.5 & M & $<50$ & B & .... & .... \\
NTT-MM25$^a$ & $4.3\pm0.6$ & 12 05 17.93 -07 43 06.9 & M & $<45$ & B & .... & .... \\
      &  & 12 05 17.59  -07 43 11.5 & P & $<45$ & B &  .... & .... \\
NTT-MM5 & $5.4\pm0.8$ & 12 05 18.15 -07 48 04.4 & M & $<45$ & B & .... & .... \\
NTT-MM1 & $5.2\pm1.0$ & 12 05 19.87 -07 49 35.8 & R & $73\pm15$ & B &  2.2 & 0.0046 \\
NTT-MM16 & $3.4\pm0.7$ & 12 05 39.47 -07 45 27.0 & P & $\rm 56\pm15$ & B & 1.9 & 0.033 \\
NTT-MM31$^b$ & $6.5\pm0.9$ & 12 05 46.54 -07 41 32.9 & R & $\rm 44\pm15$ & B & 
0.1 & $7.8\times10^{-6}$ \\
NTT-MM22 & $3.0\pm0.7$ & 12 05 43.89 -07 43 31.3 & M & $<45$ & B & .... & .... \\
\hline
A2125-MM2 & $4.2\pm0.7$ & 15 39 58.10  66 13 35.9 & R & $\rm 313\pm8$ & O & 2.1 & 0.0041 \\
A2125-MM11 & $3.1\pm0.7$ & 15 40 47.19 66 15 51.8 & R & $\rm 108\pm8$ & O & 1.4 &  0.0018 \\
A2125-MM13$^c$ & $3.6\pm0.7$ & 15 40 49.42 66 20 15.1 & R & $\rm 71\pm8$ & O & 4.5 & 0.019 \\
A2125-MM21 & $4.6\pm0.7$ & 15 41 17.85 66 22 33.7 & R & $\rm 141\pm8$ & O & 2.7 & 0.0069 \\
A2125-MM26 & $4.3\pm0.7$ & 15 41 26.9 66 14 37.3 & R & $\rm 86\pm8$ & O & 0.61 &  0.00035 \\
A2125-MM27 & $4.9\pm0.6$ & 15 41 27.29 66 16 17.0 & R & $67\pm8$ & O & 2.9 & 0.0077 \\
A2125-MM28 & $3.9\pm0.7$ & 15 41 28.78 66 22 02.7 & R & $1332\pm8$ & O & 4.5 & 0.019 \\
A2125-MM42 & $3.1\pm0.7$ & 15 42 10.66 66 21 13.0 & R & $\rm 175\pm8$ & O & 3.0 & 0.0082 \\
A2125-MM50 & $4.6\pm1.3$ & 15 42 20.27 66 07 16.0 & R & $\rm 103\pm8$ & O & 4.9 & 0.023 \\
A2125-MM32 & $4.2\pm1.0$ & 15 41 42.83 66 05 59.0 & R & $\rm 89\pm8$ & O & 0.5 &  0.00025 \\
A2125-MM29 & $3.3\pm0.7$ & 15 41 38.18 66 08 01.2 & M & $<30$ & O & .... & .... \\ \hline
\end{tabular}\)

\flushleft
(1) Galaxy name; (2) flux at 1200$\mu m$  measured in the MAMBO
survey; (3) position (RA and Dec, J2000.0) used for the
SCUBA observation; (4) provenance of position: M indicates 
a position from the original MAMBO survey, R indicates a VLA position, 
P indicates a position measured with the IRAM Plateau De Bure interferometer;
(5) radio flux at 1.4 GHz in $\mu$Jy; (6) reference for radio map:
I---Ivison et al. (2002); B---Bertoldi
et al. (2003); O---Owen et al. (2003);
(7) the distance in arcsec from the MAMBO position to the radio
position;
(8) the probability that the radio source is not associated with
the MAMBO source.\\
Notes: 
a---see Section 4 for a description of the astrometry for this
source.
b---the position given here is the radio position, which we
used for the SCUBA observation; the position
measured by the Plateau de Bure interferometer 
(12 05 46.59 -07 41 34.3) 
is only 1.6 arcsec away, much less than the size of the SCUBA
beam.
c---There are two radio sources close to the 
MAMBO position,
one 
0.8 arcsec from the MAMBO position and one 4.5 arcsec from the MAMBO position. 
The probability of either being chance associations is low (0.06\% and
1.9\%). This is not unusual for 
SMSs (Ivison et al. 2002) and is probably due to both radio 
sources being associated in some way,
possibly being in the same cluster.
The radio source
which is most likely to be physically associated with the SMS is
the closer one, and this is the position we used for our
MAMBO on-off observation (15 40 50.01, 66 20 15.1). However, by mistake
we observed the other position with SCUBA (the position
given above). 
Fortunately, the difference between the two positions 
(3.55 arcsec) is much less than the size of the SCUBA beam. 
We estimate, from the shape of the SCUBA beam at
850$\mu$m, that our 850$\mu$m measurement should be lower than
the true value by
$\simeq$20\%. In calculating the ratio
of 850$\mu$m to 1200$\mu$m flux, we have therefore increased the 850$\mu$m flux given in Table 2 by
this factor.

\end{table*}

\section{SCUBA and New MAMBO Observations}

The SCUBA submillimetre camera on the James Clerk Maxwell Telescope
is described in detail in Holland et al. (1999). It consists of
two arrays: an array of 91 bolometers for operation at short
wavelengths, usually 450$\mu$m, and an array of 37 bolometers
for use at long wavelengths, usually 850$\mu$m; a
dichroic beamsplitter
is used
to simultaneously observe the same field at the two wavelengths.
The beam size is about 8 arcsec and
14 arcsec (full-width half-maximum) at 450 and 850 $\mu$m.

We observed the MAMBO sources in photometry mode, in which a single
bolometer in each array is pointed at the target. During the commissioning
of SCUBA it was discovered that the best photometric accuracy is obtained
by averaging the signal over an area slightly larger than the beam size.
To achieve this, the secondary mirror is `jiggled' so that the bolometer
samples a 3 by 3 grid with a grid spacing of 2 arcsec. 
This was the procedure we followed. We also chopped and nodded the secondary
mirror in the standard way, using a chop throw of 60 arcsec.
Before observing the MAMBO source,
we observed a standard JCMT pointing source close to the MAMBO source
and chosen to be on the same side of the meridien, 
since the pointing and tracking accuracy of the JCMT deteriorates as the
telescope moves accross the meridien.
During each night we monitored the opacity
of the atmosphere at both 450 and 850 $\mu$m using `skydips' and
we determined the flux calibration from observations
of one or more of the standard JCMT flux calibrators.
Our observations were carried out on 14 separate nights from May 2001 until
February 2002. The dates on which the sources were observed, the
integration times, and the typical opacity of the
atmosphere during the observations are listed in Table 2.

We reduced the data in the  standard way using the SURF package
(Jenness 1997). We first subtracted the signals recorded in the
two nods from each other, 
and then divided the result by the array's flat-field.
We then corrected the result for the opacity of the atmosphere, 
interpolating
between the results from the skydips to determine
the opacity of the atmosphere at the time of the observation.
Even with chopping and nodding, and even in the best conditions,
there is usually some residual sky signal visible on the SCUBA images. 
Fortunately, this
residual signal, although variable with time, does not vary much accross the arrays,
and so the signal in the bolometers which are not pointing towards the
target can be used to removed the remaining sky signal.
We made this correction, for each second of data, by 
subtracting the median of the signals recorded by the redundant bolometers
from the signal recorded by the bolometer pointed at the target.

The result of this procedure is a plot of
intensity verses time for the target bolometer from
which all instrumental and atmospheric effects have
been removed.
Each point on the plot corresponds to one second
of data.
As the final step in the data reduction, we removed all points more
than 3$\sigma$ away from the mean and then took the mean of the remaining
points as our estimate of the intensity.

The 850$\mu$m results are listed in Table 2. The flux errors
were calculated by adding in quadrature the error obtained from the
scatter on the intensity-verses-time plot and
the photometric error
obtained from the observations of flux calibrators.
SCUBA data at 850$\mu$m is remarkably photometrically stable, and
our estimates of the photometric error were $\simeq$5-10\%.
As can be seen from the table, we observed about half the sample
on at least two different nights. Except in one case (A2125-MM26), the values measured
on the different
nights were consistent. The final
flux value for each source is a weighted average of the results
for the different nights. Our photometric measurement for LH-MM13
agrees well with that measured from a SCUBA map
by Scott et al. (2002).

At 450$\mu$m, we detected only two objects at greater than
the 3$\sigma$ level. We detected A2125-MM28 with a S/N
of 6.5 and A2125-MM11 with a S/N of 3.2. Unfortunately, because
of the greater difficulties of calibrating 450$\mu$m data
(Dunne and Eales 2001), the calibration uncertainties are
very large for these two objects. With these calibration uncertainties,
we estimate the 450$\mu$m flux of the first source
as $\rm 26\pm14$mJy and the 450$\mu$m flux of the
second source as $\rm 19\pm11$ mJy.

In addition to the SCUBA observations, we used the MAMBO array
to make new flux measurements at 1200$\mu$m of 14 of the
sources in Table 1. The aims of this were (1) to check the
fluxes obtained from the MAMBO survey and (2) by carrying out the
MAMBO photometry at the same position used for the SCUBA
photometry, to minimise the
effect of astrometric errors on our estimates of the 
850$\mu$m
to 1200$\mu m$ flux ratio. 
We used the MAMBO `on-off' mode, which is equivalent to the SCUBA
photometry mode. We reduced the data using a similar procedure to that
used
for the SCUBA photometry.
The new MAMBO
fluxes are given in the last column in Table 2.

\begin{table*}
\(\begin{tabular}{|l|l|l|l|l|l|l|l|l}
\multicolumn{9}{|c|}{Table 2. Photometry}\\

\hline
(1)&(2)&(3)&(4)&(5)&(6)&(7)&(8)&(9)\\

Name & Date & Int. time & $\rm \tau_{850\mu m}$ & $S_{850\mu m}$ & S/N &
other $S_{850\mu m}$ & $S_{1200\mu m}$/mJy & $S_{1200 \mu m}$/mJy \\
 & & (s) & & (mJy) & & (mJy) & (survey) & (on-off) \\
\hline
LH-MM13$^{\dag}$ & 20020216 & 5400 & 0.21 & $8.7\pm1.8$ & 5.7 & $10.5\pm1.6$ &  
$4.2\pm1.0$ & $3.5\pm0.5$ \\
 & 20020218 & 1260 & 0.13 & $11.0\pm2.8$ & 4.2 & ... & ... & ....  \\
 &          & & & $\bf 9.4\pm1.5$ & $\bf 7.1$  & ... & ... & ....  \\ 
LH-MM34$^{\dag}$ & & & & ... & ... & $9.5\pm2.8$ & $\rm 3.3\pm0.7$ & .... \\
LH-MM8$^{\dag}$ & & & & ... & ... & $10.9\pm2.4$ & $\rm 4.6\pm0.8$ & .... \\
LH-MM86 & 20011224 & 1800 & 0.34 & $\bf 15.0\pm3.0$ & $\bf 5.8$ & ... &  
$4.6\pm0.6$ & .... \\ \hline
NTT-MM3 & 20020207 & 1800 & 0.24 & $7.0\pm2.6$ & 2.8 & ... & $4.6\pm1.0$ & $3.2\pm1.0$ \\ 
        & 20020208 & 4770& 0.24 & $5.9\pm1.3$ & 4.9 & ...  & ... & ....\\
        &          & & & $\bf 6.1\pm1.2$ & $\bf 5.6$ & ...  & ... & .... \\
NTT-MM34 & 20020222 & 1800 & 0.25 & $1.8\pm2.3$ & 0.8 & ... & $3.3\pm0.7$ & $3.8\pm1.7$ \\
         & 20020223 & 3600& 0.22 & $-1.4\pm1.8$ & -0.8 & ... & ... & .... \\
         &          & & & $\bf -0.2\pm1.4$ &$\bf -0.1$ & ... & ... & .... \\  
NTT-MM25M$^{\dag}$ & 20020222 & 5400 & 0.24 & $5.7\pm1.5$ & $4.3$ & ... & $4.3\pm0.6$  & $3.0\pm0.5$ \\
                   & 20010514 & 3600 & 0.27 & $7.7\pm1.6$ & 5.5 & ... & ... & ....  \\
                   & 20020218 & 5400 & 0.16 & $5.3\pm1.9$ & 2.9 & ...  & ... & .... \\
                    &         & & & $\bf 6.3\pm0.9$ & $\bf 7.5$ & ... & ... & .... \\
NTT-MM25P$^{\dag}$ &20020211 & 3600 & 0.30 & $\bf 0.4\pm2.1$ & $\bf 0.2$  & ... & ... 
 & $2.7\pm0.8$ \\
NTT-MM5 & 20010521 & 6120 & 0.30 & $5.9\pm2.2$ & 2.9 & ... & $5.4\pm0.8$ & $2.2\pm1.9$ \\
         & 20020211 & 3600 & 0.36 & $7.3\pm2.0$ & 3.8 & ... & ...  & .... \\
         &          & & & $\bf 6.7\pm1.5$ & $\bf 4.8$ & ... & ... & .... \\
NTT-MM1 & 20020208 & 3600 & 0.22 & $5.5\pm1.6$ & 3.7 & ... & $5.2\pm1.0$ & $5.3\pm1.0$ \\
        & 20020209 & 3150 & 0.17 & $3.5\pm1.6$ & 2.2 & ...  & ... & .... \\
         &          & & & $\bf 4.5\pm1.2$ & $\bf 4.3$ & ... & ... & .... \\
NTT-MM16 & 20020207 & 6858 & 0.24 & $\bf 6.3\pm1.4$ &$\bf 5.3$ & ... & $3.4\pm0.7$ & $3.8\pm1.0$ \\
NTT-MM31 & 20010514 & 3600 & 0.33 & $\bf 18.5\pm2.4$ & $\bf 12.7$ & ... & $6.5\pm0.9$ & $10.6\pm1.0$ \\
NTT-MM22 & 20010522 & 5760 & 0.29 & $\bf 2.3\pm1.4$ & 
$\bf 1.6$ & ... & $3.0\pm0.7$ & ... \\
\hline
A2125-MM2 & 20020208 & 7200 & 0.23 & $\bf 5.9\pm1.3$ & $\bf 4.9$ & ... & $4.2\pm0.7$ & ... \\
A2125-MM11 & 20010522 & 6660 & & $2.1\pm1.6$ & 1.3 & ... & 
$3.1\pm0.7$ & $3.1\pm0.57$ \\ 
           & 20011223 & 3600 & 0.16 & $3.9\pm1.4$ & 2.9 & ... & ... & .... \\
           &           & & & $\bf 3.1\pm1.1$ & $\bf 3.2$ & ... & ... & .... \\
A2125-MM13$^a$ & 20020207 & 6300 & 0.30 & $4.9\pm1.6$ & 3.3 & ... & 
$3.6\pm0.7$ & $2.5\pm0.9$ \\
                     & 20020209 & 2196 & 0.21 & $8.5\pm2.6$ & 3.4 & ... & ... & .... \\
                     &          & & & $\bf 5.9\pm1.4$ & $\bf 4.7$ & ... & ... & .... \\
A2125-MM21 & 20010520 & 5580 & 0.44 & $11.7\pm2.8$ & 5.4 & ... & 
$4.6\pm0.7$ & $6.0\pm1.3$ \\
           & 20020211 & 4500 & 0.32 & $12.4\pm2.4$ & 5.9 & ... & ... & .... \\
           &          & & & $\bf 12.1\pm1.8$ & $\bf 8.0$ & ... & ... & .... \\
A2125-MM26 & 20010305 & 4320 & 0.17 & $8.5\pm1.4$ & 7.6 & ... & $4.3\pm0.7$ & $4.0\pm0.5$ \\
 & 20010514 & 2880 & 0.23 & $16.8\pm2.3$ & 10.8 & ... & ... & .... \\
 &          &      &      & $\bf 10.7\pm1.2$ & $\bf 13.2$ & ... & ... & .... \\
A2125-MM27 & 20010305 & 4320 & 0.17 & $\bf 14.6\pm1.8$ & 13.3 & ... & 
$4.9\pm0.6$ & $4.2\pm0.3$ \\
A2125-MM28 & 20011223 & 4500 & 0.14 & $\bf 5.8\pm1.2$ & $\bf 5.5$ & ... & 
$3.9\pm0.7$ & $2.8\pm0.9$ \\
A2125-MM42 & 20020222 & 2160 & 0.36  & $0.4\pm2.5$ & 0.2 & ... & $3.1\pm0.7$ & .... \\
           & 20020223 & 3240& 0.19 & $5.8\pm2.1$ & 2.9 & ... & ... & .... \\
           &          & & & $\bf 3.6\pm1.6$ & $\bf 2.9$ & ... & ... & .... \\
A2125-MM50 & 20020223 & 5400 & 0.18 & $\bf 6.5\pm1.6$ & $\bf 4.3$ & ... & $4.6\pm1.3$ & ... \\
A2125-MM32 & 20020222 & 5400 & 0.24 & $\bf 6.00\pm1.6$ & $\bf 4.1$ & ... & $4.2\pm1.0$ & ... \\
A2125-MM29 & 20020211 & 3600 & 0.33 & $\bf -2.6\pm2.2$ & $\bf -1.2$ & ... & $3.3\pm0.7$ & ... \\ \hline
\end{tabular}\)
\flushleft
(1) Galaxy name. A dagger indicates
there is a note on this source in Section 4; 
(2) date on which the observation was made; 
(3) integration time in seconds---this includes the
time spent at the reference position;
(4) optical depth of the atmosphere
at 850$\mu$m during the observation;
(5) flux at 850$\mu$m in mJy measured from this observation. The
measurement in bold type is the weighted average of the
individual measurements for this
source.
(6) signal-to-noise of observation;
(7) flux measurement at 850$\mu$m for this
object from Scott et
al. (2002);
(8) flux measurement at 1200$\mu m$ in mJy from
MAMBO survey;
(9) flux measurement at 1200$\mu m$ in mJy from subsequent MAMBO on-off
observations.
Notes: a---see note on this source in Table 1.

\end{table*}

\section{Notes on Individual Sources}

In this section we give notes, when appropriate, on individual sources:

\noindent {\bf LH-MM13:} There is a large offset (7.6 arcsec) between the
positions of the 
MAMBO source and the VLA radio source, much the largest for our sample of
MAMBO sources (\S 5.2), although the statistical analysis shows 
that the probability this is a chance association is only 6\%. This source 
is in the region mapped with SCUBA at 850$\mu$m 
by Scott et al. (2002). The brightest source in their survey, L850.1, is much closer (3.0 arcsec) 
to the radio source. This positional disagreement has recently been resolved
by a new MAMBO map, which shows the
position of the MAMBO source is quite close to the radio and SCUBA positions.
The initial disagreement is thought to have been caused by the effect
of a large value for the noise close to the edge of the original MAMBO 
map (Bertoldi, private communication).
The position we used for our 
SCUBA and MAMBO photometry is that of the radio source. 
Our 850$\mu$m measurement and that of Scott et al. 
agree well. 

\noindent {\bf LH-MM34} and
{\bf LH-MM8:} We did not observe these MAMBO sources with SCUBA, but they
fall in the
region mapped with SCUBA by Scott et al. (2002). 
The positions of the SCUBA sources L850.14 and 
L850.2, whose fluxes we 
give in Table 2, agree well with 
the positions of the two MAMBO sources.

\noindent {\bf NTT-MM25:} The positions given in Table 1 are the original MAMBO
position and the position of the source detected with the 
Plateau de Bure
interferometer at 1260$\mu$m by Dannebauer et
al. (2002). The positions are separated by 6.8 arcsec. The obvious inference to
draw from this would be that the original MAMBO position was substantially in error.
However, we detected much stronger 850$\mu$m emission at the original MAMBO position
than at the position measured by the interferometer. 
Recent new data taken with the interferometer, when added to the original data, has
produced a new position for the SMS (12 05 17.86, -07 43 08.5), which
is only 1.9 arcsec away from the original survey position (Bertoldi, private
communication).
When calculating the ratio of 
850$\mu$m and 1200$\mu$m flux for this source, we 
used the SCUBA and MAMBO on-off measurement at the original survey position.

\section{A Comparison of the SCUBA and MAMBO Results}

A basic result of our SCUBA observations is that most
of the 4$\sigma$ sources detected by the MAMBO survey
are real. Of the 23 MAMBO sources for which there are SCUBA observations,
only four were not detected at $\rm >3\sigma$ with SCUBA. Of these
four, one was detected by SCUBA at 2.9$\sigma$ and one other
was detected by the VLA. Therefore, of the 23 MAMBO sources, there
are only two for which there is not corroborating
evidence in another waveband. In this section, we will use our new datasets
to investigate some of the basic characteristics of
the SCUBA and MAMBO surveys. We will first use our new MAMBO
photometry to investigate the issue of flux-boosting. We will then use
the radio and MAMBO datasets to investigate the astrometric
accuracy of the MAMBO survey.
Finally, we will start
to compare the fluxes measured for individual SMSs by MAMBO and by SCUBA.

\subsection{Flux Boosting}

Several groups (Hogg 2001; Eales et al. 2000; Scott et al. 2002) have argued
that `flux-boosting' will be important in submillimetre and
millimetre surveys.
Flux boosting is an effect which may occur in any magnitude- or flux-limited
sample of sources. If the differential source counts
decrease with flux density, a source in the final sample is more likely to
have had its flux density enhanced than depressed by the effect of noise.
Flux boosting is a particular example of 
the process described by Eddington (1940), in which the statistical
properties of a 
a distribution of experimental measurements are distorted by noise.
In the MAMBO and SCUBA surveys, there are two components
to the noise: the noise arising from the experimental setup (instrumental
and atmospheric) and the confusion noise arising from the SMSs which
are too faint to be detected individually.
Eales et al. (2000) 
carried out a Monte-Carlo simulation 
of their 850$\mu$m SCUBA survey, concluding
that the average boosting factor is $\simeq$44\%.
Scott et al. (2002) carried out a Monte-Carlo simulation
of their
`8 mJy' SCUBA survey, concluding 
that for this survey the boost factor is $\simeq$15\%.

We can address the question of flux boosting in the MAMBO survey empirically 
using our new MAMBO photometry.
If an SMS was boosted into the original MAMBO survey
by noise, the new flux should be lower
than the original survey flux. 
This is strictly true only for the first type of noise; as long
as the direction and size of the chop throw
used in the photometry and the original
survey were the same, 
the two fluxes
will have been 
affected by the confusion of faint sources in the same way.
We can therefore 
only set a lower limit on the flux boosting factor. 
The positions from the MAMBO
survey will, of 
course, have been affected by both instrumental noise and the confusion
of faint sources, but since our photometric observations 
were almost always made at
the accurate radio positions, this should not be a serious issue.

Figure 1 shows 
the ratio of 
the 1200$\mu$m fluxes measured from the survey and from the
photometry plotted against the survey flux.
We have also plotted the
results of a Monte-Carlo simulation
of the MAMBO survey (\S 7). Inspection of
Table 2 shows that there is only one SMS for which the MAMBO survey 
flux and the MAMBO photometric flux are significantly different, and the
general
good agreement 
is evidence that fluxes measured with MAMBO are
reproducible. In the figure, 
the effect of flux boosting can be clearly seen in the
Monte-Carlo simulation, but is less clear in the real data.
Excluding the one SMS with a big difference between the two fluxes
and one SMS which was detected with low signal-to-noise in the
photometry, the
average ratio of the survey flux density
to the photometric flux density is $\rm 1.14\pm0.07$. 
This is slightly lower than the flux-boosting factors inferred for
the SCUBA surveys, although our estimate for the MAMBO survey
is strictly only
a lower limit. The difference might also be explained by the smaller
beam of the MAMBO survey (11 arcsec) compared with the SCUBA
surveys (14 arcsec).

\begin{figure}
\psfig{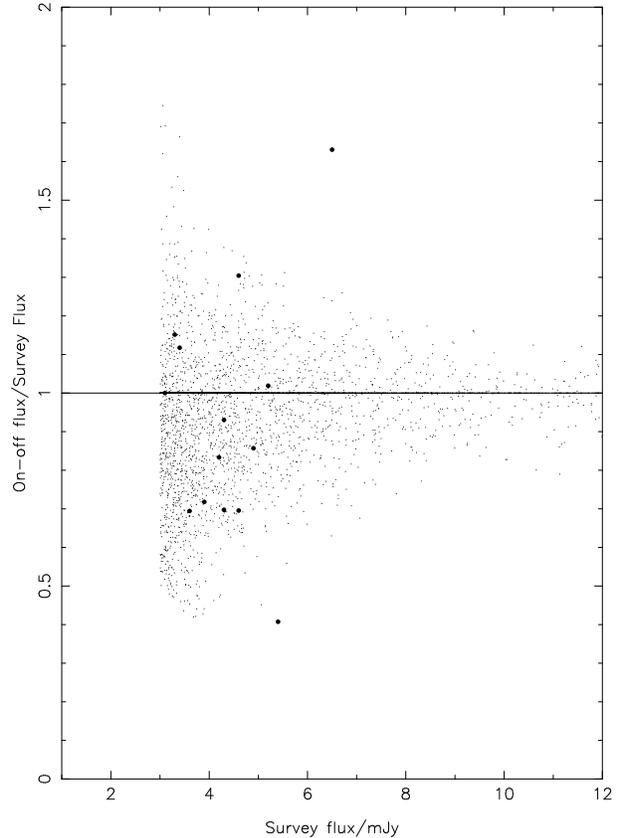}
\caption{The large symbols show the 
ratio of the 1200$\mu$m flux measured using the
MAMBO photometry (`on-off') mode to the flux measured
by the survey, plotted against the survey flux.
The small
points show the results from the Monte-Carlo simulation
of the MAMBO survey described in \S 7. For these points the
quantity plotted on the y-axis is the
ratio of the
1200$\mu$m flux 
in the absence of noise to the flux after
noise has been added. The quantity plotted on the x-axis is
the flux after noise has been added.
Only sources which would have been detected
in the survey have been plotted.
}
\end{figure}

\subsection{The accuracy of the MAMBO positions}

Many authors have tried to quantify the positional accuracy of
submillimetre and millimetre surveys using Monte-Carlo simulations
(e.g. Eales et al. 2000; Hogg 2001; Scott et al. 2002).
We can address this question empirically for the MAMBO
survey using the differences
between the positons from the survey and the accurate positions
(measured either with the VLA or the Plateau de Bure interferometer).
Figure 2 shows a histogram of these differences. The median difference
is 2.0 arcsec, fairly modest given the size of the MAMBO
beam (FWHM of 11 arcsec). Nevertheless, there are a few SMSs
with large offsets, suggesting that the effect of source confusion
may sometimes be important.

\begin{figure}
\psfig{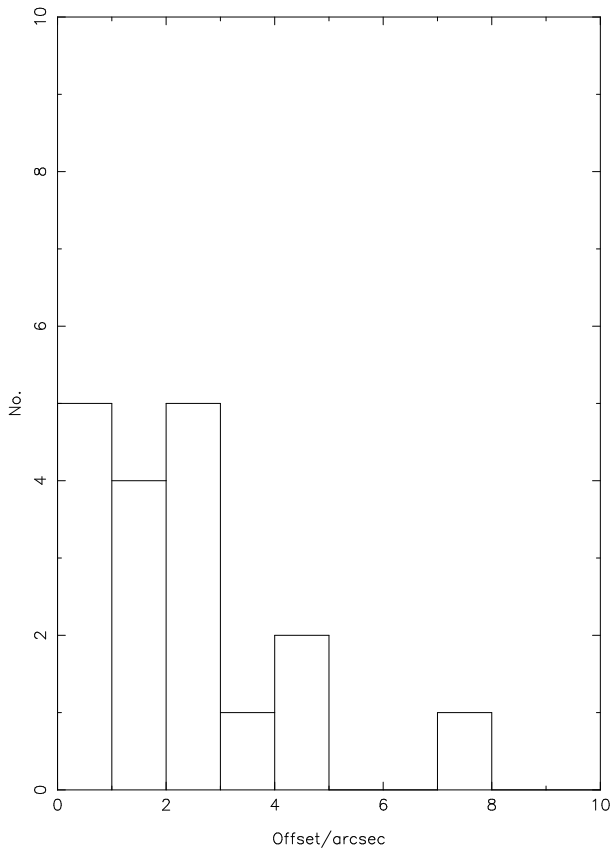}
\caption{Difference between the position from the MAMBO survey
and the position measured with the VLA or the Plateau de Bure interferometer.
}
\end{figure}

\subsection{A Comparison of the MAMBO and SCUBA Fluxes}

The ratios of
850$\mu$m and 1200$\mu$m flux for the SMSs in our sample are much lower
than expected for a low-redshift galaxy.
The explanation
of this difference may either be astrophysical (the SMSs are either
at higher redshifts or have different rest-frame spectral energy
distributions to galaxies at
low redshift) or there might be some systematic error
as the result of our experimental method.
In this section we will consider
the latter possibility.

An obvious possibility is that there is an error in the
absolute flux calibration of one of the telescopes.
However, 
the primary flux calibrators of both telescopes are the same
objects: the planets Mars, Uranus and Neptune. Moreover, the sets
of
secondary calibrators for both telescopes (used when a planet
is not visible) have a large overlap (Sandell 1994; Lisenfeld et al.
2000). Lisenfeld et al. have compared the fluxes of 11 secondary
calibrators measured at a similar wavelength with the IRAM 30-metre
and
the JCMT, finding that the ratio of JCMT-to-IRAM flux has an average
value of 0.99 with a standard deviation of 0.13. Thus, it seems certain that
the low values of the 850 to 1200 $\mu$m flux ratio are not
caused by an error in the absolute flux calibration.

A second effect to consider is the effect of
bandwidth. The filter used in the SCUBA observations
has a central frequency of 347 GHz and a bandwidth (FWHM) of 30 GHz.
The spectral
response of MAMBO is more complicated. Carilli et al. (2001) give
the half-power sensitivity range as 210-290 GHz but note that the
overall profile is asymetric, with a sharp rise in sensitivity
at lower frequency, and then a gradual decrease in sensitivity to
higher frequency. The instrumental spectral responses of both instruments
should, in principle, be convolved with the transmission curve of
the atmosphere, which will depend on the conditions in which a particular
observation was made. We have made an estimate of the effect of the finite
bandwidth by assuming that the spectral response for both instruments
has the form of a top hat, with the width of the top hat being the measured
FWHM. We have assumed that an SMS has a power-law spectral
energy distribution ($S_{\nu} \propto \nu^{\alpha}$).
For values of $\alpha$ of 1, 2, 3 and 4, our estimates
of the values measured for the flux ratio through these filters
are 1.39, 1.91, 2.61, and 3.54, respectively.
Our estimates of the values that would be 
measured 
through filters of zero width are 1.39, 1.93, 2.67, and
3.71, respectively. 
Thus, it is only
when the flux ratio is very large that the effect of bandwidth
becomes important, and the measured values of the flux
ratio of the
MAMBO sources
are generally much less than this.

The final issue we need to address are the effect of astrometric errors
on the SCUBA 850$\mu$m photometry.
Two types of astrometric errors might be important: (1) 
an error in the position assumed
for the MAMBO source, (2) JCMT pointing/tracking errors.
We can eliminate the first possibility easily, because for
all but five of the 23 sources
there is either an accurate position (from the VLA or the Plateau de Bure
interferometer) or the 
850$\mu$m flux was measured from a submillimetre
map. For the remaining five sources we were forced to
use the MAMBO survey position but,
as we showed in the previous section, these positions are quite accurate.
The median error in Figure 2 (2 arcsec) is much less than the beam size of
SCUBA (FWHM of 14 arcsec),
and
a
pointing error of this size would lead to 
an underestimate of the true 
850$\mu$m flux by only 6\%.

We can investigate the effect of the second type of error using our
JCMT dataset. We always observed a JCMT pointing source before, and
in most cases after, observing
the MAMBO source (\S 3).
The difference between the actual and 
expected position of the second pointing source can be used
to investigate the effects of tracking and pointing. Figure 3
shows a histogram of this difference for all observations
in which the second pointing source was in the same part of the
sky as the target. The median discrepancy is 1.91 arcsec.  

\begin{figure}
\psfig{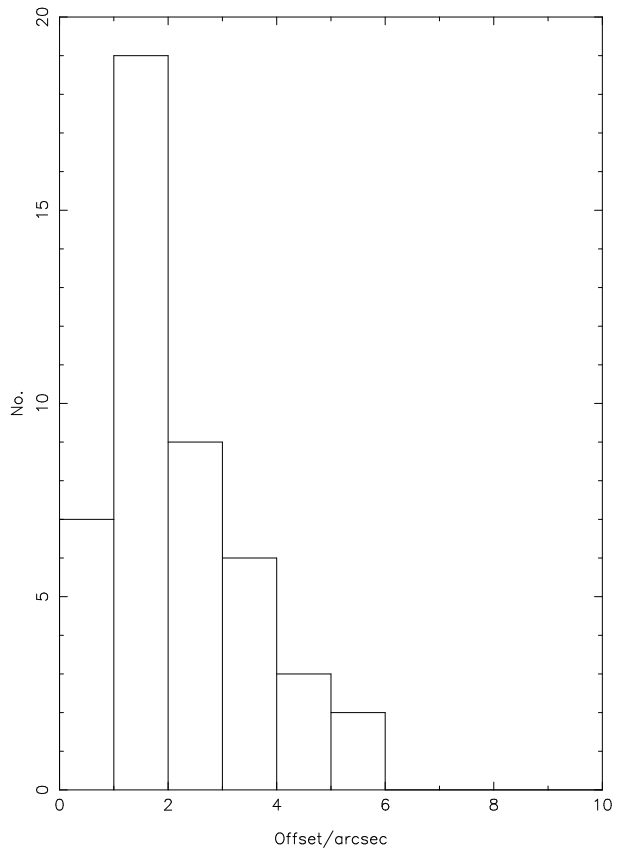}
\caption{Difference between the actual and expected
positions of the JCMT pointing
source observed after the observation of the MAMBO source.}
\end{figure}

The way this is interpreted depends on whether tracking or pointing
errors are most important. Let us first suppose
the positional discrepancies are caused
by tracking errors. On the assumption that
the tracking errors accumulate linearly with time, the average positional error
during an observation will be about half that shown in the figure.
The median positional error due to tracking problems is therefore 
0.96 arcsec, which would lead to an underestimate of the true
850$\mu$m flux by $\simeq$1\%.
Now let us suppose that the discrepancies are caused by pointing errors.
On the assumption that the pointing error on moving to the target and
the pointing error on moving to the second pointing source add in quadrature,
the median pointing error would be $1.91/\sqrt{2} = 1.35$ arcsec. An error
of this size
would lead to an underestimate of the 850$\mu$m flux by $\simeq$3\%.

Therefore, we conclude that astrometric errors, of whichever kind, should
have a relatively small effect on the 850$\mu$m fluxes. Furthermore,
in calculating the ratio of the 850$\mu$m and 1200$\mu$m flux of an
SMS we have used,
where possible, the 1200$\mu$m flux measured from our follow-up
photometry. Since this photometry was carried out at the same position
and should have the same pointing/tracking concerns as the SCUBA photometry,
for these SMSs there is no reason at all to expect astrometric errors to lead
to an underestimate of the
flux ratio.

\section{Astrophysical Explanations of the Low Flux Ratios}

Figure 4 shows a comparison of the 850/1200$\mu$m
flux ratios of the MAMBO sources
with the flux ratios of dust sources of known redshift. 
The spectral energy distributions of high-redshift
SMSs are very poorly known. In a search through
the literature, we were unable to find
a single high-redshift SMS without an obvious
active nucleus which had a sufficiently well-sampled
spectral energy distribution (SED) for a dust temperature
to be determined. Some SMSs have been observed at
several submillimetre wavelengths (Dey et al. 1999; Ivison
et al. 2000), but not at a short
enough wavelength in the rest-frame to sample properly the
peak of the SED, the crucial part of the SED for determining the
temperature of the dust.
Our set of comparison dust sources
in the figure are a sample of quasars
observed both with MAMBO 
and with SCUBA (Omont et al. 2001; Isaak et al. 2002), a combined
SCUBA-MAMBO dataset very similar to ours
(an additional check that calibration differences are not
an issue). 
The quasars are mostly at $\rm z \sim 4$. We do not know the redshifts
of the MAMBO sources, but if they are like the SCUBA sources
observed by Chapman et al. (2003) they will lie in the redshift
range $\rm 0.8 < z < 4$. In the figure we have plotted them at an
arbitrary redshift of 0.2.

The values of the flux ratio
for the MAMBO sources are generally lower than those for the
quasars. There are two alternative explanations for this.
As we show in detail below, the value of this flux ratio
is expected to decline with redshift. Therefore, one possible
explanation for the difference between the MAMBO sources and
the quasars is that the MAMBO sources are generally at higher redshifts
than the quasars.
The alternative is that the rest-frame SEDs of the MAMBO sources
are different than those of the quasars.

We have plotted in the figure the predicted relationship between
the 850/1200$\mu$m
flux ratio and redshift for various SEDs. 
The SED of a dust source is usually represented as
\smallskip
$$
S_{\nu} = \nu^{\beta} \sum_i a_i B_{\nu}(T_i)
$$
\smallskip
\noindent in which $S_{\nu}$ is the
flux density at a frequency $\nu$, $B_{\nu}$
is the Planck function and $\beta$ is the dust emissivity index.
The sum is over the components of dust 
at different temperatures, with $a_i$ being the relative
mass of the dust at a temperature $T_i$.
The first set of predictions we have plotted in the
figure are for a simple model in which there is only one
dust component in a galaxy. We have plotted predictions based
on this model, for a number of temperatures and for two values
of $\beta$.

The 850/1200$\mu$m flux ratios of
the quasars are well-fitted
by the predictions for SEDs with a single dust temperature
and a value for $\beta$ of 2. This conclusion is supported
by more detailed investigations of the SEDs of high-redshift
quasars. Priddey and McMahon (2002) find that the
SEDs of high-redshift quasars can be well represented
by dust at a single temperature with $\rm \beta=2$;
Benford et al. (1999) find that the SEDs are fitted well
by single-temperature dust with $\rm \beta=1.5$.
If the
rest-frame SEDs of the MAMBO sources 
can also be represented by dust at a single temperature with a high
value of $\beta$, it is clear from the diagram that the MAMBO sources
must generally be at higher redshifts than the quasars, and outside the
range found by Chapman et al. (2003) for SCUBA sources.

However, observations of luminous low-redshift dust sources
do not reveal SEDs like this. 
Dunne et al. (2000) and Blain, Barnard and Chapman (2003) have shown 
that the 850$\mu$m and IRAS far-infrared fluxes of low-redshift galaxies
can be represented by dust at a single temperature. Dunne and Eales
(2001---henceforth DE), however, have 
shown that when 450$\mu$m fluxes are added, it
is often no longer possible to fit the fluxes with a single-temperature
model. Furthermore, they have shown that  
the ratio
of 
450 and 850$\mu$m fluxes is remarkably constant over a wide
range of galaxies types, from normal-looking spiral galaxies to
ULIRGs such as Arp 220. The constancy and high value of this
ratio
require $\beta$ to be close to two. With this
value of $\beta$ it is
impossible to fit the fluxes with
dust at a single temperature for virtually all
galaxies. DE conclude 
that there are at least two dust
components in nearby galaxies, with most of the dust being at
$\rm T_d \sim 20K$, even in the case of ULIRGSs like Arp 220.
Amure (2003) has reached the same conclusion with an independent dataset
containing 15 spiral galaxies. Note, however, that this conclusion
does not imply that all dust in a galaxy has a value for $\beta$ of
2 or that there are only two dust components in a galaxy. Observations
within the Galaxy have shown that the properties of dust depend strongly
on environment (Dupac et al. 2002; Stepnik et al. 2003) and in
reality there must be a range of dust temperatures. The
DE conclusion is an empirical one: to fit the global
submillimetre and far-infrared fluxes of nearby galaxies, it is necessary
to use at least two dust components and a global value for $\beta$ of 2.
Priddey and McMahon (2002) have also reached the conclusion that
the rest-frame SEDs of high-redshift quasars are different from those
of low-redshift galaxies.

We estimated redshifts for the MAMBO sources in the following
way, starting with the assumption that, in the rest frame,
they
are like the low-redshift galaxies modelled by DE.
We first fitted the DE model to the submillimetre
and far-infrared data for 104 galaxies in the IRAS Bright Galaxy
Survey (DE; Dunne et
al. 2000).
We then used the model to predict how the 850/1200$\mu$m ratio
should depend on redshift for each galaxy, taking proper
account of the effects of the microwave background (Eales
and Edmunds 1996).
The thick line in Figure 4 shows the median predicted flux ratio
at each redshift
for the 104 galaxies, with the dashed lines showing the
lowest and highest prediction.

We considered each MAMBO source in turn, determining the
redshift at which each DE template produced the observed
value
of the 850/1200$\mu$m flux ratio.
This produced 104 redshift estimates for each MAMBO
source.
We took the median of these redshift estimates
as our estimate of the redshift for the MAMBO source.
There are two sources of error on this estimate: (1) an error
arising from the uncertainties in the 850 and 1200
$\mu$m fluxes, (2)
an error arising
from fact that a different template might be a better
representation of the rest-frame SED of the MAMBO source.
We estimated these errors 
separately. 
We estimated the first error using the DE
template which had
given the median redshift estimate. 
We allowed the redshift to vary in both directions until 
the Chi-square agreement between the predicted and observed
fluxes (one degree of freedom) was sufficiently poor
that the probability of it occurring by chance
was 16\%, effectively placing 
$\pm1\sigma$ errors
on the redshift estimate. We estimated the error 
from possibly using the wrong template in the
following way.
We ranked the 104 redshift estimates.
The redshift 
estimate below which 16\% of the estimates fell, and 
the redshift estimate above which 16\% of the estimates
fell, provided $\pm$1$\sigma$ errors. 
We are, of course, making the
assumption that the 104 templates represent
the full range of SEDs of high-redshift SMSs.
We added the two errors in quadrature. Table 3 
lists the redshift estimates and the errors. 
In several cases we have given a lower or upper limit
to the redshift. This is for galaxies where the observed
value of the flux ratio fell either below the value
predicted by the median template at all redshifts
or above the value
predicted by the median template at all redshifts. 
The limits are $\pm1\sigma$ limits,
in the sense that there is a probability of 16\% that the
true redshift is actually either below the lower limit 
given in the table or above the upper limit given in the table.

The errors on the redshift estimates are 
very large, mostly caused by the errors in the flux densities rather 
than by the diversity of possible templates, but in general the
redshift estimates are
very high. Fifteen out of 21 MAMBO sources for which there is corroborating
evidence in other wavebands (\S 5) have redshifts estimated by this
method of $>$3.

Is there any obvious astrophysical way of avoiding this conclusion?
One way to do this would be to alter the templates used to
estimate the redshifts. If we assumed a single-temperature template
with a value for $\beta$ of 2, which reproduces the flux ratios
of the quasars, it would actually make things worse, leading
to even higher redshifts.
A way to lower the redshift estimates, however, would be
to use a single-temperature template with a value for
$\beta$ of 1.
Figure 4 shows that the
low values of the flux ratio might be explained by a low
value for $\beta$ with the MAMBO
sources lying in the redshift range found by Chapman et al. (2003).
If this is the case, the global properties of the dust in
high-redshift SMSs must be different from those of dust in low-redshift
galaxies.

\begin{figure}
\psfig{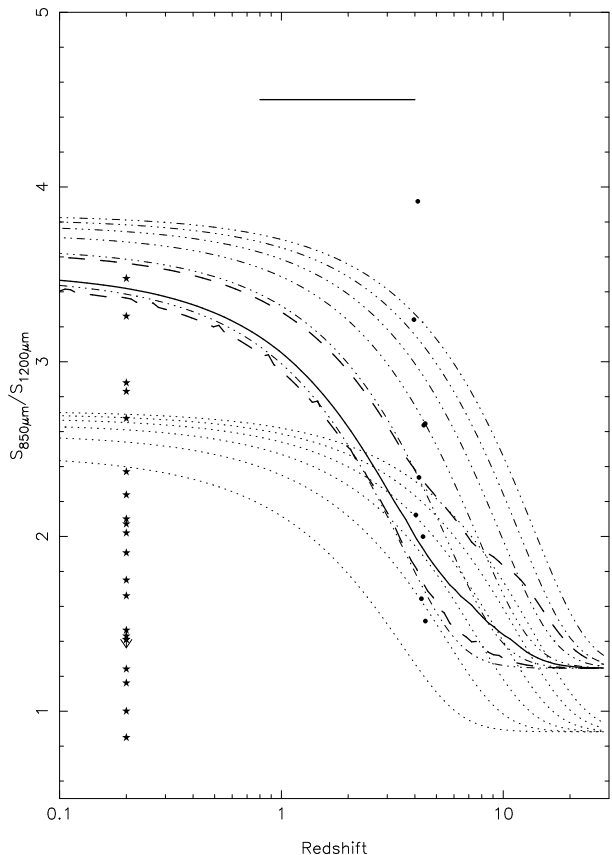}
\caption{The 
ratio of $\rm S_{850 \mu m}$ and $\rm S_{1200 \mu m}$ flux verses
redshift. The asterisks show the measured values of the
flux ratio for the MAMBO sources (we omit the two
sources for which there is no corroborating evidence
in another waveband---\S 5). Since we do not know the redshifts
of the MAMBO sources, we have plotted the points at an arbitrary
redshift of 0.2. 
The dots show the values of the flux ratio
for a sample of high-redshift
quasars (Omont et
al. 2001; Isaak et al. 2002). The horizontal
line shows the range of spectroscopic
redshifts for SCUBA sources
found by Chapman et al. (2003). The thin lines show
the relationship between flux ratio and redshift predicted
for a single-temperature dust model, the dotted
lines for $\rm \beta = 1$, the dot-dashed lines
for $\rm \beta = 2$. For both sets, the lowest and highest
lines are for
dust temperatures of 20K and 70K, respectively, 
with the other lines
being at an interval of 10K.
The thick lines are predictions based on the models
of Dunne and Eales (2001---see text). 
The continuous line shows the median 
predicted value of the flux ratio and the dashed lines show the
lowest and highest predicted value at each redshift.
}
\end{figure}

In the absence of spectroscopic redshifts, the only
other information about the redshifts of the MAMBO
sources comes from the
the ratio
of radio and submillimetre flux, which, as Carilli and Yun (1999)
originally pointed out, should be a function of redshift.
Following the original suggestion, a
number of groups 
used different samples of low-redshift objects to 
determine the expected relationship between 
the ratio of radio-to-submillimetre flux
and redshift
(Carilli and Yun 2000; Dunne, Clements and
Eales 2000; 
Rengarajan and Takeuchi 2001). There 
are slight differences between the redshifts estimated
using the different sets of low-redshift templates
(Ivison et al. 2002), but for
our work these differences are not important.
We estimated a redshift for each MAMBO source using the
observed ratio of radio-to-submillimetre flux or the limit
on this ratio if the MAMBO source was not detected by the VLA.
For low-redshift templates we used the
25 most
radio-luminous sources from the sample of Dunne, Clements
and Eales (2000), which produces very similar redshift estimates
to using the templates in
Carilli and Yun (2000).
We used exactly the same method for estimating the redshift
as we used to
estimate the redshift from
the 850/1200$\mu$m flux ratio. As before, we estimated separately
the errors arising from flux errors and 
from the range of possible templates, and then added the errors
in quadrature. 
We are of course assuming again that the templates
represent the full range of SEDs of high-redshift SMSs.
The results are shown in Table 3.

There are two obvious differences in the two sets of redshift
estimates. First, the
errors on the new set of redshift estimates are much smaller,
mainly because the ratio of radio-to-submillimetre
flux is a stronger function of redshift
than the 850/1200$\mu$m flux ratio. Second, the redshifts estimated from the
radio-to-submillimetre ratio are in general lower than those from the
850/1200$\mu$m flux ratio. Let us consider the number of sources
with estimated redshifts $\rm > 3$. Using the radio method, there are
six out of 21 sources with estimated redshifts $\rm >3$, if we include sources
which have redshift lower limits below three as being above this redshift.
If we use the other method, there are 15 sources with estimated redshifts
$\rm z >3$ (we assume that NTT-MM34 is
at $\rm z >3$ because although we could not obtain a satisfactory fit
to its 850$\mu$m and 1200$\mu$m fluxes, the very low value
for the 850/1200$\mu$m ratio suggests a very high redshift).
Therefore, although in the cases of individual sources the 
large errors on the redshift
estimates often mean that the two redshift estimates are statistically
consistent, there is a trend for the redshifts estimated using the
radio method to be lower.

Is there any reason to conclude that one of the two methods
is biased? 
We have no independent check on the redshifts estimated
from the 850/1200$\mu$m flux ratio because there
are virtually no SMSs with 
measurements of this flux ratio and spectroscopic redshifts. 
However, we can 
critically examine the second technique because there
are now a significant number of SMSs with radio measurements
and spectroscopic redshifts. Figure 5 
shows the ratio of submillimetre to radio flux plotted against
redshift for all SMSs which have
both spectroscopic redshifts and
radio measurements. Out of the 21 SMSs, 14 have positions which
fall close to the positions predicted using low-redshift
templates.
Seven out of 21, however, fall a significant
way from the predicted positions.
The most interesting SMSs, from the point of view of this
work, are the four which have much higher
ratios of radio to submillimetre flux than the predicted values.
One possible explanation of the excess radio emission is that these
SMSs contain radio-emitting active nuclei. Therefore,
one way to reconcile the two sets
of redshift estimates in Table 3 would be if 
our
MAMBO sample contains
a large number of SMSs with radio-emitting active nuclei.
The alternative way to reconcile the two sets of redshift
estimates
is to conclude that the radio estimates are
correct and that the low values for the 850/1200$\mu$m flux ratio
are caused by the rest-frame SEDs of the MAMBO sources being different
from those of low-redshift galaxies.

To summarize the results of this section: The low 850$\mu$m/1200$\mu$m
flux ratios of our sources may be explained by very high redshifts,
although the redshifts estimated using the radio method are much
lower. If the radio estimates are correct, the most likely explanation
of the low flux ratios is that the global properties of dust
in the MAMBO sources are different from those of dust in local galaxies.
There is recent evidence that the global dust emissivity index in
local galaxies ($\beta$) is close to two. If $\beta$ is closer to
one in high-redshift SMSs, this would be a natural explanation
of the low values of the flux ratio.

\begin{figure}
\psfig{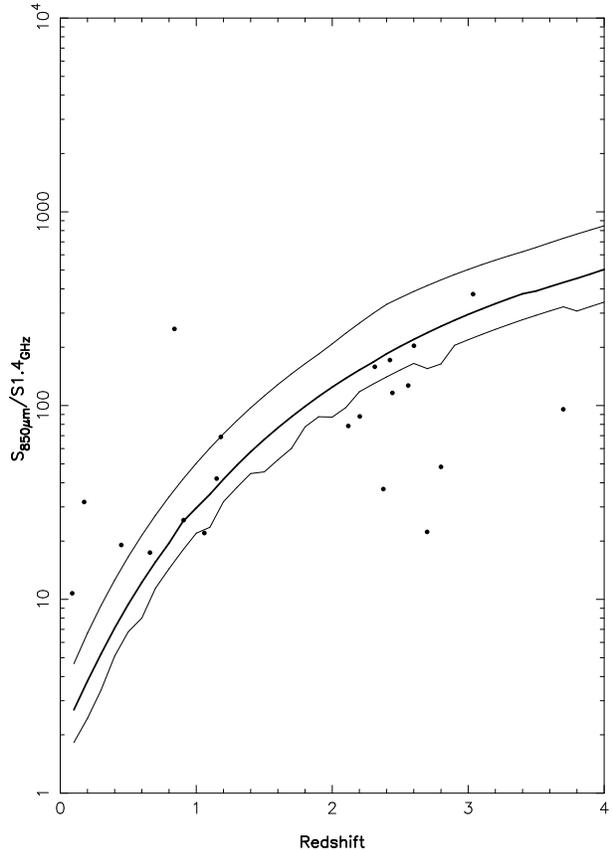}
\caption{The ratio of
850$\mu$m flux to 1.4-GHz flux verses redshift.
The lines show predictions 
based on the sample of low-redshift star-forming
galaxies of Dunne et al. (2000). 
We used the SED
of each galaxy to predict how the submillimetre to radio
ratio should depend
on redshift. The thick line shows the median prediction at each redshift,
the thin lines show quasi-$\pm1\sigma$ predictions based on the range
of predicted values at each redshift (see Dunne, Clements and Eales 2000).
The points show SMSs with spectroscopic redshifts and radio detections.
The data are from Eales et al. (2000), 
Smail et al. (2000), Ivison et al. (2002), Chapman et al. (2002,
2003), Simpson et al.
(2003), Clements et al. (2003).
}
\end{figure}

\begin{table*}
\begin{tabular}{|l|l|l}
\multicolumn{3}{|c|}{Table 3. Redshift Estimates}\\
\hline
(1)&(2)&(3)\\

Name & $z_{est}$ & $z_{est}$  \\
 & radio & $1200 \mu m / 850 \mu m$  \\  
\hline

LH-MM13 & $2.35_{-0.64}^{+0.71}$ & $3.05_{-3.05}^{+5.72}$ \\
LH-MM34 & $2.85_{-0.86}^{+1.0}$ & $1.35_{-1.35}^{+5.61}$ \\
LH-MM8 & $3.45_{-0.78}^{+1.61}$ & $2.65_{-2.65}^{+6.11}$ \\
LH-MM86 & $2.55_{-0.45}^{+0.72}$ & $0.45_{-0.45}^{+2.60}$\\
\hline
NTT-MM3 & $2.05_{-0.42}^{+0.67}$ & $4.25_{-4.25}^{+\infty}$ \\
NTT-MM34$^a$ & $>2.25$ & ..... \\
NTT-MM25 & $>2.85$ & $3.65_{-2.53}^{+4.26}$ \\
NTT-MM5 & $>3.15$ & $12.75_{-8.63}^{+\infty}$\\
NTT-MM1 & $2.45_{-0.67}^{+0.64}$ & $>8.95$ \\
NTT-MM16 &  $2.25_{-0.64}^{+0.85}$ & $6.35_{-4.85}^{+\infty}$ \\
NTT-MM31 & $3.45_{-1.03}^{+1.3}$ & $5.25_{-1.93}^{+6.14}$ \\
\hline
A2125-MM2 & $1.25_{-0.22}^{+0.51}$ & $10.35_{-6.04}^{+\infty}$\\
A2125-MM11 & $1.75_{-0.36}^{+0.54}$ & $>9.65$ \\
A2125-MM13 & $2.05_{-0.72}^{+0.64}$ & $1.55_{-1.55}^{+5.11}$ \\
A2125-MM21 & $2.45_{-0.67}^{+0.67}$ & $3.95_{-2.83}^{+8.44}$ \\ 
A2125-MM26 & $2.25_{-0.51}^{+0.63}$ & $1.85_{-1.61}^{+1.92}$ \\
A2125-MM27 & $2.45_{-0.51}^{+0.63}$ & $<1.0$ \\
A2125-MM28 & $0.55_{-0.28}^{+0.22}$ & $3.45_{-3.45}^{+\infty}$ \\
A2125-MM42 & $1.55_{-0.36}^{+0.51}$ & $>3.65$\\
A2125-MM50 & $3.15_{-1.0}^{+0.85}$ & $10.15_{-8.10}^{+\infty}$ \\
A2125-MM32 & $2.25_{-0.57}^{+0.67}$ & $9.85_{-7.53}^{+\infty}$ \\

\hline
\end{tabular}
\flushleft

(1) Galaxy name. A letter as a superscript
indicates
there is a note on this source below;
(2) redshift estimate from the ratio of the 1.4-GHz and 1200$\mu$m 
fluxes using the method described in the text; 
(3) redshift estimate from the 
850$\mu$m and $1200 \mu m$ fluxes, using the method described
in the text.\\
Notes:
a---the absence of a redshift estimate indicates that we could
not obtain a satisfactory fit to the data at any redshift 
with any of the templates.

\end{table*}

\section{The cosmic evolution of dust---a statistical approach}

In this section 
we assess the statistical evidence that the low values of the
850/1200$\mu$m flux ratio
imply there is a significant population of SMSs at very high redshift
(We assume that the low values are due to high redshifts
rather than to a difference in the properties of dust,
an assumption which may of course be wrong).
Since the errors on the redshift estimates for the individual sources
are so large, it is necessary to consider the sample as a whole.
We have adopted the Monte-Carlo approach of generating artificial
samples of sources on the assumption that there are no high-redshift
SMSs, and then testing this assumption by comparing the 
flux ratios of the artificial samples
with the flux ratios of the real sample.
The Monte-Carlo approach allows us to include the effect
on our artificial samples of the noise in the SCUBA and MAMBO
observations.

In this Monte-Carlo simulation, we have made the standard assumption that the
luminosity function of SMSs
can be factorised into its
dependence on luminosity and on redshift: $\Phi(L,z) = E(z)\phi(L)$.
We have to make four assumptions: (1) an assumption about the
form of $E(z)$;
(2) an assumption about the form of $\phi(L)$; (3) a cosmological model; 
(4) an assumption about the SEDs of SMSs. 
The
first assumption, of course, is what we are really trying to test.
Here, as we are only interested in whether
there is any evidence for a high-redshift population
of SMSs, we can make some simplifying assumptions.
We have first assumed that there are no SMSs
at $\rm z < 1$. This is demonstrably not true
(e.g. Chapman et al. 2003; Webb et al. 2003), but the vast majority of SMSs
do appear to be beyond this redshift. The assumption is also
a conservative one because without it
the artificial
samples would contain more high values of
the $\rm S_{850 \mu m}/S_{1200 \mu m}$ flux ratio. Above this redshift,
we have assumed that $E(z)$ is independent of redshift up to a maximum
redshift, $z_{max}$. By comparing the observed flux ratios with the
artificial flux ratios for different values of $z_{max}$, we can address
the question of whether there is a population of high-redshift SMSs.

The number of sources
expected in a sample of sources with fluxes above a flux limit,
$S_{min}$, as a function of redshift (the selection
function) is given by:
\smallskip
$$
n(z) = E(z) \int^{\infty}_{L(S_{min},z)} \phi(L) dL {dV \over dz}.
$$
\smallskip
\noindent 
In this equation $V$ is the comoving volume. $L(S_{min},z)$ is the
minimum luminosity an SMS could have at a redshift $z$ and still be detected
in the sample. This quantity is very weakly dependent on redshift, 
and so the precise form of $\phi(L)$ has little effect on $n(z)$.
We assumed that
$\phi(L)$ has the same power-law form as the differential source
counts: $\phi(L) \propto L^{-\alpha}$.
We used 2.6 as our standard
value of $\alpha$, which was the value
found by Bertoldi et al. (2001) for the 1200$\mu$m source counts.
We tried different values of $\alpha$ but, as expected, it made almost
no difference to the final results.

It is also necessary to make an
assumption about the typical SED of an SMS, in order to
calculate the lower limit of the integral. 
We used two
extreme
SEDs from the sample of IRAS galaxies of Dunne et al.
(2000). NGC 958 is a galaxy whose SED
is dominated by cold dust.
The observed fluxes of this galaxy are fitted well
by the two-component dust model of Dunne and Eales (2001), with dust
at 20K and 44K in the ratio by mass of 186:1. At the other
extreme is the galaxy IR1525$+$36, which, in the Dunne and Eales model, has
dust at 19K and 45K in the ratio by mass of 15:1. 

The first step in the simulation was 
to generate 40000 SMSs. The equation above is effectively a redshift
probability distribution, and we used
a random
generator to produce a redshift for each SMS. 
We then used
the luminosity function (the same used
in the calculation of the
selection
function) and a random number generator
to produce a luminosity for the source.
From the luminosity and redshift of each SMS and the cosmological
model we calculated the
flux of the SMS.
We produced SMSs with fluxes well below those which would have been detected
in the MAMBO survey to allow for the possibility of noise boosting the
SMS into the survey. We carried out a separate simulation for
each of the two SEDs.

The next step was to add the effects of observational noise. We first
used a random number generator to add on gaussian noise to the
1200$\mu$m flux of each SMS, with a level similar to
the noise in the real MAMBO survey. 
After eliminating the SMSs which would have fallen below the flux limit
of the real MAMBO survey,
we produced 1000 samples of SMSs,
each of 21 sources. We then used the same SED we had used
to calculate the selection function to calculate the
850$\mu$m source of each SMS, and then used a random number generator
to add on gaussian noise similar to the noise of the SCUBA photometry.

Figure 6 shows the values of the
$\rm S_{850 \mu m}/S_{1200 \mu m}$ flux ratio for
the real sample and the distributions 
predicted by two simulations, both with
$\rm z_{max} = 3$. In one simulation we have used the 
hot SED and in one the cold SED. 
The observed distribution
appears quite different from the results of both
simulations, a difference which is statistically significant
(one-sample KS test) 
at $<<$1\% level in both cases. 
Therefore, if the low values of the 
$\rm S_{850 \mu m}/S_{1200 \mu m}$ flux ratio
are caused by the effect of redshift, there is strong statistical
evidence that the MAMBO samples contain a significant number of
SMSs at $\rm z > 3$.

\begin{figure*}
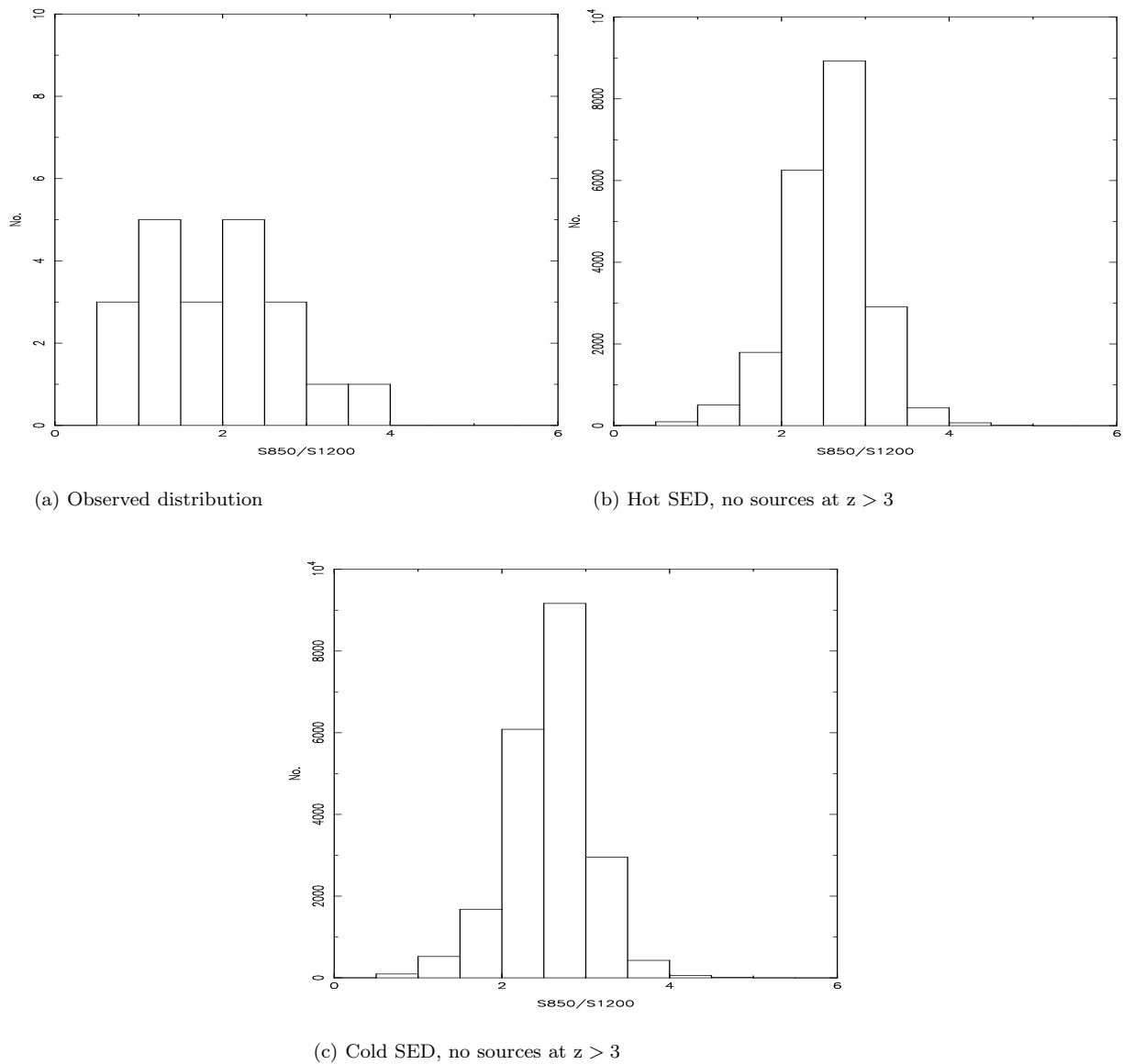

\subfigure[Observed distribution]{\psfig{file=Figure6a.ps,height=6.5cm,width=8.0cm}}
\subfigure[Hot SED, no sources at $\rm z > 3$]{\psfig{file=Figure6b.ps,height=6.5cm,width=8.0cm}}
\subfigure[Cold SED, no sources at $\rm z > 3$]{\psfig{file=Figure6c.ps,height=6.5cm,width=8.0cm}}
\caption{A comparison of the observed and
predicted distributions of the $S_{850\mu m}/S_{1200 \mu
m}$ flux ratio. Figure 6(a) shows the distribution for the
real sample.
The other two figures show the predictions of
models in which there are
no SMSs at $\rm z > 3$. In both models we have 
assumed the `concordance universe'
with
$\rm \Omega_M = 0.3$, $\rm \Omega_{\Lambda} = 0.7$. In Figure 6(b) we have
assumed the hot SED, in Figure 6(c) we have assumed
the cold SED.}
\end{figure*}

\section{Discussion}

The two possible explanations of our observational results
are either that the dust emissivity index is low in the
SMSs or that the SMSs are at extremely high redshifts.
Both possibilities have interesting implications.

Although extragalctic submillimetre astronomers usually make the
convenient assumption that the emissivity of dust is a universal
constant (James et al. 2002), there is now plenty of 
evidence that it varies
from place to place even within the Galaxy (e.g. Stepnik et al. 2003).
As well as evidence for the
variation of the emissivity at a particular wavelength, there is also
evidence that the dust emissivity index varies. 
Dupac et al. (2002), for
example, have used data taken with the balloon-borne submillimetre
telescope PRONAOS to show that the dust emissivity index appears to
depend inversely on dust temperature, decreasing to a value close
to one in the hot parts of star-formation regions. Moreover, models
of the evolution of dust in
prestellar cores predict that the emissivity index should
decrease in regions of increased density (Ossenkopf and Henning 1994). 
Therefore, although there is evidence that the dust emissivity index
for the emission from galaxies as a whole is close to two at
low redshifts (Dunne
and Eales 2001; Amure 2003), it would not be suprising,
in the possibly quite different conditions in the interstellar medium of
SMSs, if the emissivity index were close to one.

The alternative explanation is that a significant fraction of the
MAMBO sources lie at very high redshifts. There are two interesting
points here. First, if SMSs are ellipticals being seen during the
formation process, our result is more in line with the traditional
model in which ellipticals form at early times in a distinct
`epoch of galaxy formation' (e.g. Larson 1975), rather than the
current paradigm that ellipticals form by hierarchical merging
over a long period of cosmic time. The second interesting point
is the constraint on the process that forms the dust. Some of
our redshift estimates are as high as ten, when the
time since the Big Bang is only 0.42 Gyr in the concordance
model ($\Omega_M = 0.3$, $\Omega_{\Lambda}=0.7$).
Given the shortness of time, it is difficult to see how the dust
in these objects could have been made in the atmospheres of
evolved stars (Morgan and Edmunds 2003), and one
is forced to assume that supernovae are the source
of the dust. 

\section{Conclusions}

We have observed 23 sources from the MAMBO 1200$\mu$m survey with
SCUBA at 850$\mu$m, detecting 19 of the sources.
The sources generally have low values for the ratio of
850$\mu$m to 1200$\mu$m flux. 
At face value, the low values of the flux ratio
imply very high redshifts, with 15 of the MAMBO sources
having estimated redshifts $\rm >3$. Redshifts estimated using
the ratio of radio to submillimetre flux are, however, much lower.
If the latter estimates are correct, the most likely explanation
of the low values of the 850/1200$\mu$m flux ratio is that
the global properties of the dust in high-redshift SMSs are different
from those of dust in local galaxies.
There is recent evidence that the global dust emissivity index in
local galaxies ($\beta$) is close to two. If $\beta$ is closer to
one in high-redshift SMSs, this would be a natural explanation
of the low values of the flux ratio.

\section*{Acknowledgments}

Stephen Eales thanks the Leverhulme Trust for the award of a research
fellowship. We thank Phil Mauskopf, Haley Morgan and Robert Priddey
for useful conversations and Andrew Blain for many useful comments
on the paper.
The Institute for Radioastronomy at Millimeter Wavelengths (IRAM) is
funded by the German Max-Planck-Society, the French CNRS, and the Spanish
National Geographical Institute.
The James Clerk Maxwell Telescope is operated
by the Joint Astronomy
Center
on behalf of the UK Particle Physics and
Astronomy Research Council, the Netherlands Organization for Scientific
Research and the Canadian National Research Council.
The National
Radio Astronomy Observatory (NRAO) is operated by Associated
Universities, Inc. under a cooperative agreement with the National
Science Foundation.

{}


\begin{thebibliography}{}

\bibitem[\protect\citeauthoryear{Almaini}{2003}]{}
Almaini, O. et al. 2003, MNRAS, 339, 397.

\bibitem[\protect\citeauthoryear{Amure}{2003}]{}
Amure, M. 2003, Ph.D. thesis, Cardiff University.

\bibitem[\protect\citeauthoryear{Barger}{1998}]{}
Barger, A.J. et al. 1998, Nature, 394, 428.

\bibitem[\protect\citeauthoryear{Benford}{1999}]{}
Benford, D.J., Cox, P., Omont, A., Phillips, T. \&
McMahon, R. 1999, ApJ, 518, L65.

\bibitem[\protect\citeauthoryear{Bertoldi}{2000}]{}
Bertoldi, F. et al. 2000, A \& A, 360, 92.

\bibitem[\protect\citeauthoryear{Bertoldi}{2001}]{}
Bertoldi, F., Menten, K.M. Kreysa, E., Carilli, C.L. \& Owen, F.
2001, 24th meeting of the IAU, Joint Discussion 9, Manchester, England
(astro-ph 0010553).

\bibitem[\protect\citeauthoryear{Bertoldi}{2003}]{}
Bertoldi, F. et al. 2003, in preparation.

\bibitem[\protect\citeauthoryear{Blain}{1999}]{}
Blain, A.W., Kneib, J.-P., Ivison, R.J. \& Smail,
I. 1999, MNRAS, 512, 87.

\bibitem[\protect\citeauthoryear{Blain}{1999}]{}
Blain, A.W. 1999, MNRAS, 309, 955.

\bibitem[\protect\citeauthoryear{Blain}{2003}]{}
Blain, A.W., Barnard, V.E. \& Chapman, S.C. 2003,
MNRAS, 338, 733.

\bibitem[\protect\citeauthoryear{Carilli}{1999}]{}
Carilli, C.L. \& Yun, M.S. 1999,
ApJ, 513, L13.

\bibitem[\protect\citeauthoryear{Carilli}{2000}]{}
Carilli, C.L. \& Yun, M.S. 2000, ApJ, 530, 618.

\bibitem[\protect\citeauthoryear{Carilli}{2001}]{}
Carilli, C.L. et al. 2001, ApJ, 555, 625.

\bibitem[\protect\citeauthoryear{Chapman}{2002}]{}
Chapman, S.C., Smail, I., Ivison, R.J., Helou, G.,
Dale, D.A. \& Lagache, G. 2002, ApJ, 573, 66.

\bibitem[\protect\citeauthoryear{Chapman}{2003}]{}
Chapman, S.C. et al. 2003, Nature, in press.

\bibitem[\protect\citeauthoryear{Clements}{2003}]{}
Clements, D. et al., in preparation.

\bibitem[\protect\citeauthoryear{Dannebauer}{2002}]{}
Dannebauer, H., Lehnert, M.D., Lutz, D., Tacconi, L., Bertoldi, F.,
Carilli, C., Genzel, R. \& Menten, K. 2002, ApJ, 573, 473.

\bibitem[\protect\citeauthoryear{Dey}{1999}]{}
Dey, A., Graham, J.R., Ivison, R.J., Smail, I., Wright,
G.S. \& Liu, M.C. 1999, ApJ, 519, 610.

\bibitem[\protect\citeauthoryear{Dunne}{2000}]{}
Dunne, L., Clements, D. \& Eales, S. 2000, MNRAS, 319, 813.

\bibitem[\protect\citeauthoryear{Dunne}{2000}]{}
Dunne, L., Eales, S., Edmunds, M, Ivison, R., Alexander, P.
\& Clements, D. 2000, MNRAS, 315, 115.

\bibitem[\protect\citeauthoryear{Dunne}{2001}]{}
Dunne, L. \& Eales, S. 2001, MNRAS, 327, 697.

\bibitem[\protect\citeauthoryear{Dunne}{2003}]{}
Dunne, L., Eales, S. \& Edmunds, M. 2003, MNRAS, in press
(astro-ph 0210260).

\bibitem[\protect\citeauthoryear{Dupac}{2002}]{}
Dupac, X. et al. 2002, A \& A, 392, 691.

\bibitem[\protect\citeauthoryear{Eales}{1999}]{}
Eales, S.A., Lilly, S., Gear, W., Dunne,
L., Bond, J.R., Hammer, F., Le F\`evre, O. \& Crampton, D. 1999,
ApJ, 515, 518.

\bibitem[\protect\citeauthoryear{Eales}{2000}]{}
Eales, S.A., Lilly, S., Webb, T., Dunne, L., Gear, W.,
Clements, D. \& Yun, M. 2000, AJ, 120, 2244.

\bibitem[\protect\citeauthoryear{Eales}{1996}]{}
Eales, S.A. \& Edmunds, M. 1996, MNRAS, 280, 1167.

\bibitem[\protect\citeauthoryear{Eddington}{1940}]{}
Eddington, A.S. 1940, MNRAS, 100, 354.

\bibitem[\protect\citeauthoryear{Efstathiou}{2003}]{}
Efstathiou, A. \& Rowan-Robinson, M. 2003, MNRAS, in press.

\bibitem[\protect\citeauthoryear{Hogg}{2001}]{}
Hogg, D.W. 2001, AJ, 121, 1207.

\bibitem[\protect\citeauthoryear{Holland}{1999}]{}
Holland, W.S. et al. 1999,
MNRAS, 303, 659.

\bibitem[\protect\citeauthoryear{Hughes}{1998}]{}
Hughes, D.H. et al. 1998, Nature, 394, 241.

\bibitem[\protect\citeauthoryear{Isaak}{2002}]{}
Isaak, K.G., Priddey, R.S., McMahon, R.G., Omont, A.,
Peroux, C., Sharp, R.G. \& Withington, S. 2002, MNRAS,
329, 141.

\bibitem[\protect\citeauthoryear{Ivison}{2000}]{}
Ivison, R. et al. 2000, MNRAS, 315, 209.

\bibitem[\protect\citeauthoryear{Ivison}{2002}]{}
Ivison, R.J. et al. 2002, MNRAS, 337, 11.

\bibitem[\protect\citeauthoryear{James}{2002}]{}
James, A., Dunne, L., Eales, S. \& Edmunds, M.G. 2002,
MNRAS, 335, 753.

\bibitem[\protect\citeauthoryear{Jenness}{1997}]{}
Jenness, T. 1997, SURF - SCUBA user reduction
facility. {\it Starlink User Note 216.1}.

\bibitem[\protect\citeauthoryear{Kreysa}{1998}]{}
Kreysa, E. et al. 1998, Proc. SPIE Vol. 3357, Advanced
Technology MMW, Radio, and Terahertz Telescopes,
ed. T.G. Phillips, p319-325.

\bibitem[\protect\citeauthoryear{Larson}{1975}]{}
Larson, R.B. 1975, MNRAS, 173, 671.

\bibitem[\protect\citeauthoryear{Lisenfeld}{2000}]{}
Lisenfeld, U., Thum, C., Neri, R. \& Sievers, A. 2000,
Memo on IRAM web site (www.iram.fr).

\bibitem[\protect\citeauthoryear{Omont}{2001}]{}
Omont, A., Cox, P., Bertoldi, F., McMahon, R.G.,
Carilli, C. \& Isaak, K.G. 2001, A \& A, 374, 371.

\bibitem[\protect\citeauthoryear{Ossenkopf}{1994}]{}
Ossenkopf, V. and Henning, T. 1994, A \& A, 291, 943.

\bibitem[\protect\citeauthoryear{Owen}{2003}]{}
Owen, F. et al. 2003, in preparation.

\bibitem[\protect\citeauthoryear{Morgan}{2003}]{}
Morgan, H. \& Edmunds, M. 2003, MNRAS, in press.

\bibitem[\protect\citeauthoryear{Priddey}{2002}]{}
Priddey, R.S. \& McMahon, R.G. 2002, MNRAS, 324, 17p.

\bibitem[\protect\citeauthoryear{Rengarajan}{2001}]{}
Rengarajan, T.N. \& Takeuchi, T. 2001, PASJ, 53, 433.


\bibitem[\protect\citeauthoryear{Richards}{2000}]{}
Richards, E.A. 2000, ApJ, 533, 611.

\bibitem[\protect\citeauthoryear{Sandell 1994}{1994}]{}
Sandell, G. 1994, MNRAS, 271, 75.

\bibitem[\protect\citeauthoryear{Scott}{2002}]{}
Scott, S. et al. 2002, MNRAS, 331, 817.

\bibitem[\protect\citeauthoryear{Smail}{1997}]{}
Smail, I., Ivison, R.J. and
Blain, A.W. 1997, ApJ, 490, L5.

\bibitem[\protect\citeauthoryear{Smail}{2000}]{}
Smail, I., Ivison, R.J., Owen, F.N.
Blain, A.W. \& Kneib, J.-P. 2000, ApJ, 528, 612.

\bibitem[\protect\citeauthoryear{Smail}{2002}]{}
Smail, I., Ivison, R.J., Blain, A.W. \& Kneib, J.-P.
2002, MNRAS, 331, 495.

\bibitem[\protect\citeauthoryear{Smail}{2003}]{}
Smail, I., Chapman, S.C., Ivison, R.J., Blain, A.W.,
Takata, T., Heckman, T.M., Dunlop, J.S. \&
Sekiguchi, K. 2003, submitted to MNRAS (astro-ph
0303128).

\bibitem[\protect\citeauthoryear{Simpson}{2003}]{}
Simpson, C. et al., in preparation.

\bibitem[\protect\citeauthoryear{Stepnik}{2003}]{}
Stepnik, B. et al. 2003, A \& A, 398, 551.

\bibitem[\protect\citeauthoryear{Webb}{2003}]{}
Webb, T. et al. 2003, ApJ, 587, 41.

\bibitem[\protect\citeauthoryear{Waskett}{2003}]{}
Waskett, T. et al. 2003, MNRAS, in press (astro-ph 0301610).


\end{thebibliography}
\end{document}